# First-principles study of defects and doping limits in CaO

Zhenkun Yuan and Geoffroy Hautier*

*Thayer School of Engineering, Dartmouth College, Hanover, New Hampshire 03755, USA*

Email: geoffroy.hautier@dartmouth.edu

**ABSTRACT**

Calcium oxide (CaO) is a promising host for quantum defects because of its ultrawide band gap and potential for long spin coherence times. Using hybrid functional calculations, we investigate the intrinsic point defects and how they limit Fermi-level positions and doping in CaO. Our results reveal calcium and oxygen vacancies to be the most common intrinsic defects, acting as compensating acceptors and donors, respectively. Oxygen interstitials are also prevailing under O-rich conditions and act as compensating donors. Due to compensation by these defects, O-poor conditions are required to dope CaO *n*-type, while O-rich conditions are required for *p*-type doping. We find that, at room temperature, intrinsic CaO can only achieve Fermi-level positions between 1.76 eV above the valence-band maximum (VBM) and 1.73 eV below the conduction-band minimum (CBM). If suitable shallow dopants can be found, the allowed range of Fermi levels would increase to between VBM+0.53 eV and CBM−0.27 eV and is set by the compensating intrinsic defects. Additionally, we study hydrogen impurities, and show that hydrogen will limit *p*-type doping but can also act as shallow donor when substituting oxygen ($H_O$ defects).

## I. Introduction

Wide-band-gap materials are attractive hosts for solid-state quantum defects as they present substantial opportunities for accommodating paramagnetic color centers that may be used in quantum information science.[1-5] Prime examples of these hosts are diamond and silicon carbide, with the nitrogen-vacancy (NV) center in diamond being one of the most important quantum defects.[6] In the past few years, a large variety of wide-band-gap oxides, nitrides, and chalcogenides have emerged as alternative hosts.[4,7] Among them, the alkaline-earth metal oxides (BeO, MgO, and CaO) are naturally of interest because of their ultrawide band gaps which could accommodate well separated defect levels.[8-12]

Calcium oxide (CaO) is especially of interest. Besides having an ultrawide band gap (7.09 eV), this material is featured by a nuclear-spin-dilute crystal lattice and therefore holds great promise for obtaining long electron spin coherence times ($T_2$).[9,13] Kanai *et al.* revealed that CaO possesses a computed $T_2$ of 34 ms, longer than those in most known materials and compared to ~1 ms in natural diamond and silicon carbide.[9] These results have motivated recent efforts to explore quantum defects in CaO, and a class of NV-like centers in CaO has recently been computationally identified.[12] In wide-band-gap materials, the Fermi level can however be challenging to control and defect compensation effects often pin the Fermi level, limiting what charged defects can be stabilized.[14-18] This calls for clarifying what range of Fermi levels could be achieved in CaO. Specifically, the most promising one of the recently predicted NV-like centers requires to be made in a moderately to highly *n*-type doped CaO with the Fermi level close to the conduction-band edge.[12] It is unclear if that would be even possible.

Here we investigate the intrinsic point defects in CaO using first-principles hybrid-functional calculations. All the intrinsic point defects are considered, including the calcium vacancy ($V_{\text{Ca}}$), oxygen vacancy ($V_{\text{O}}$), calcium interstitial ($\text{Ca}_i$), oxygen interstitial ($\text{O}_i$), calcium-on-oxygen antisite ($\text{Ca}_{\text{O}}$), and oxygen-on-calcium antisite ($\text{O}_{\text{Ca}}$). We focus on their role as charge-compensating centers, and assess how they limit Fermi-level positions and doping. It is found that despite compensation by the intrinsic defects, mainly $V_{\text{Ca}}$ and $V_{\text{O}}$ as well as $\text{O}_i$, CaO can have a wide attainable Fermi-level range of VBM+0.53 eV to CBM−0.27 eV. Additionally, we study hydrogen impurities in CaO. Hydrogen is a common impurity, and is known to cause unintentional doping and compensation in wide-band-gap materials.[19-23] Our results show that, if present, hydrogen impurities would cause additional limit to *p*-type doping, but on the other hand, hydrogen substitution on the oxygen site ($\text{H}_{\text{O}}$) is shallow donor which could contribute to *n*-type doping.

## II. Computational details

Our first-principles defect calculations were performed using the projector augmented-wave (PAW) pseudopotential method and hybrid density functional of Heyd–Scuseria–Ernzerhof (HSE) as implemented in the VASP code.[24-26] Ca 3*s*, 3*p*, and 4*s*, O 2*s* and 2*p*, and H 1*s* were treated as valence electrons. For the CaO conventional unit cell, an energy cutoff of 500 eV was used for the plane-wave basis set, and a $8 \times 8 \times 8$ $\Gamma$-centered **k**-point grid was used for Brillouin-zone integration. The HSE mixing parameter ($\alpha$) was set to 0.498, which reproduces the experimental band gap (7.09 eV[27]) and yields lattice constant (4.78 Å) close to experiment (4.78 Å,[28] 4.808 Å[29]). The point defects were simulated using a 512-atom supercell, which is a $4 \times 4 \times 4$ repetition of the HSE-relaxed CaO conventional unit cell. For the defect calculations, a $\Gamma$-only **k**-point grid and an energy cutoff of 400 eV were used. All internal atomic positions in supercells containing a point defect were fully relaxed until the residual forces became less than 0.01 eV/Å; local defect geometries are provided in the supplementary material. Spin polarization was explicitly considered in all the defect calculations.

The defect formation energies depend on the Fermi level and elemental chemical potentials in the standard formalism [see Eq. S(1) of the supplementary material]. Based on the HSE-calculated CaO formation enthalpy, $\Delta H_\text{f}(\text{CaO}) = -6.23$ eV, the oxygen chemical potential ($\mu_\text{O}$) can vary from $-6.23$ to $0$ eV, reflecting CaO growth conditions that can vary from (extreme) O-poor to O-rich. For hydrogen impurities, we chose hydrogen chemical potential ($\mu_\text{H}$) to correspond to equilibrium with $H_2$ molecules under O-poor conditions and with $Ca(OH)_2$ under O-rich conditions. More details are provided in the supplementary material.

## III. Results and discussion

### A. Intrinsic point defects

Figs. 1(a)–1(b) show the calculated formation energies of the intrinsic defects. Due to the ultrawide band gap of CaO, the defect formation energies can change significantly as the Fermi level moves over the energy gap. Under O-poor conditions [Fig. 1(a)], the defects with the lowest formation energies are $V_\text{O}$ and $V_\text{Ca}$. For $V_\text{Ca}$, this holds for Fermi-level positions close to the CBM. Under O-rich conditions [Fig. 1(b)], the $O_i$ is also a low-energy defect. Other intrinsic defects, including $Ca_i$, $Ca_\text{O}$, and $O_\text{Ca}$, have high formation energy. Since we will focus on the role of the intrinsic defects as charge-compensating centers, in the following we will limit our discussion to the $V_\text{Ca}$, $V_\text{O}$, and $O_i$.

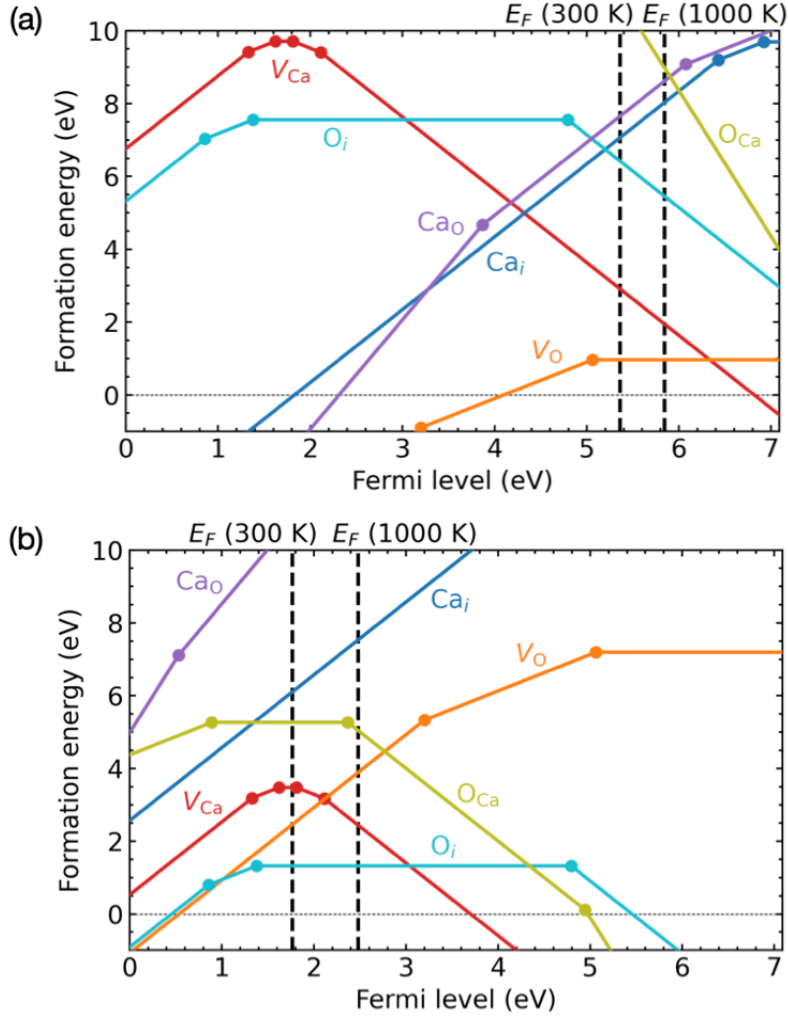

FIG. 1. Formation energy versus Fermi level for intrinsic point defects in CaO, under (a) O-poor and (b) O-rich conditions. The Fermi level is referenced to the VBM of CaO. The slopes of the formation-energy lines indicate the defect charge states, i.e., $V_{Ca}$: 2+, +, 0, −, 2−; $V_O$: 2+, +, 0; $Ca_i$: 2+, +, 0; $O_i$: 2+, +, 0, 2−, with the dots denoting the charge-state transition levels. The equilibrium Fermi levels (without extrinsic doping) at $T = 1000$ K and those quenched to 300 K are indicated by the vertical dashed lines.

We find that the $V_{Ca}$ is an amphoteric defect, exhibiting both positive and negative charge states across the CaO band gap. It behaves as a deep acceptor when the Fermi level is above the (0/−) transition level, at 1.81 eV above the valence-band maximum (VBM). The findings are different from previous semilocal functional calculations which failed to stabilize the positive charge states of $V_{Ca}$ and showed that $V_{Ca}$ acts exclusively as a shallow acceptor.[30] The $V_O$ is a deep double donor, similar to the behavior in other alkaline-earth oxides.[10, 31] The $O_i$ is amphoteric; its neutral charge state is stable over a wide range of the band gap, and the positive and negative charge states occur for Fermi levels close to the VBM and CBM, respectively.

Using the formation energies and assuming the absence of any impurities, we have determined the equilibrium Fermi level at $T = 1000$ K (which is within the typical range of temperatures for electrical measurements on CaO samples[32-35]); see supplementary material for more details. Under O-poor conditions, the equilibrium Fermi-level position is pinned by $V_O^+$ and $V_{Ca}^{2-}$ and lies at 1.25 eV below the CBM [see Fig. 1(a)]. This leads to a net electron concentration on the order of $10^{14}$ cm$^{-3}$, a relatively low $n$-type doping level. By contrast, under O-rich conditions, the equilibrium Fermi-level position is pinned by $O_i^+$ and $V_{Ca}^{2-}$ and lies at 2.48 eV above the VBM [see Fig. 1(b)]. This gives a net hole concentration on the order of $10^8$ cm$^{-3}$, an extremely low *p*-type doping level (probably not measurable). Since we have been considering the O-poor and O-rich limits, the equilibrium Fermi-level positions for other chemical-potential conditions will be deeper in the band gap, corresponding to lower doping levels. Our results agree with previous electrical measurements that as-grown CaO samples are very poor conductors even at high temperatures.[32-35] Our work also explains the experimentally observed oxygen partial-pressure dependence of the electrical conductivity and its type (*n*-type and *p*-type) in CaO.[34, 35]

In addition, we have considered rapid quenching from 1000 to 300 K, assuming that the concentration of each of the defect species is fixed to that at 1000 K (but the concentration of different charge states will redistribute at 300 K).[36-39] As shown by the dashed lines in Fig. 1, after quenching, for O-poor CaO, the Fermi level lies further below the CBM (1.73 eV below the CBM), while for O-rich CaO, the Fermi level lies closer to the VBM (1.76 eV above the VBM). Thus, assuming high-temperature synthesis followed by rapid quenching to room temperature, we expect the Fermi level of intrinsic CaO to range from VBM+1.76 eV to CBM−1.73 eV.

### B. Doping limits

We now focus on the role of the intrinsic defects as charge-compensating centers, and quantify their limits to doping based on the calculated formation energies. The basic mechanism is that as the Fermi level is driven towards the band edges by extrinsic doping or electrostatic gating, the formation energy of intrinsic compensating defects becomes lowered leading to increased compensation.[14, 40] The Fermi level at which intrinsic compensating acceptors (donors) start to have negative formation energy and hence form spontaneously is the "pinning energy" for *n*-type (*p*-type) doping.[14, 40-43] The energy range where one can shift the Fermi level is between the *n*-type and *p*-type pinning energies. The ability to dope a material depends on the position of these pinning energies with respect to the band edges.

From the above, we see that the $V_{Ca}$ acts as the dominant compensating acceptor, and that the $V_O$ and $O_i$ act as the dominant compensating donors. As seen in Fig. 1(a), under extreme O-poor conditions, the *p*-type and *n*-type pinning energies are at VBM + 4.10 eV and CBM − 0.27 eV, set by $V_O^+$ and $V_{Ca}^{2-}$, respectively. The allowed range of Fermi levels is thus restricted to the upper half of the band gap. Since the *p*-type pinning energy is well above the VBM, *p*-type doping is unlikely, while for *n*-type doping, Fermi-level positions within 0.27 eV of the CBM will not be attainable. Under extreme O-rich conditions [Fig. 1(b)], the *p*-type and *n*-type pinning energies are at VBM + 0.53 eV and CBM − 3.39 eV, set by $V_O^{2+}$ and $V_{Ca}^{2-}$, respectively, so that the allowed range of Fermi levels is restricted mostly to the lower half of the band gap. Under this condition, *n*-type doping is unlikely, while for *p*-type doping, Fermi-level positions within 0.53 eV of the VBM will not be attainable.

Extending the discussion to all chemical-potential conditions, in Fig. 2 we plot the pinning energies as a function of $\mu_O$. We see that as $\mu_O$ varies from −6.23 eV (the O-poor limit) to 0 eV (the O-rich limit), both the *p*-type and *n*-type pinning energies are consistently lowered, and correspondingly, the allowed range of Fermi levels shifts from the upper half to lower half of the band gap. Irrespective of $\mu_O$, the *n*-type pinning energy is always defined by the $V_{Ca}^{2-}$. By contrast, the *p*-type pinning energy is defined by $V_O^+$ for $\mu_O$ below −5.36 eV and by $V_O^{2+}$ for

$\mu_O$ above this value. Fig. 2 clearly shows that O-poor conditions are required to dope CaO $n$-type, while O-rich conditions are required for $p$-type doping.

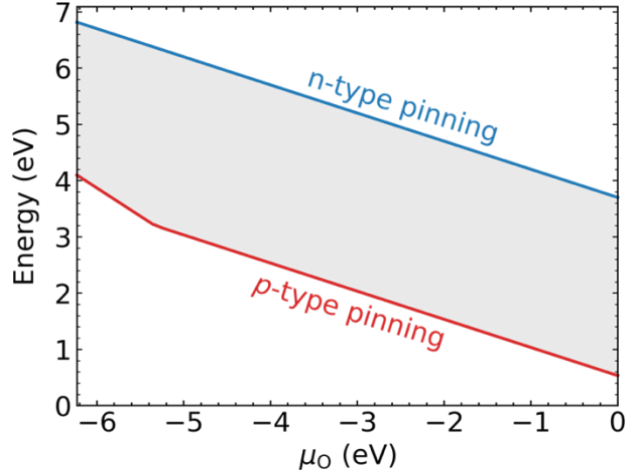

FIG. 2. Doping limits in CaO as a function of oxygen chemical potential ($\mu_O$), considering the intrinsic defect compensation. The VBM is set as the energy zero (y-axis). The $p$-type and $n$-type pinning limits (denoted by the red and blue lines, respectively) define the allowed range of Fermi levels (shaded area).

Overall, CaO has a doping-limit energy range of VBM+0.53 eV to CBM−0.27 eV, which spans about 90% of the CaO band gap. Such a large allowed range of Fermi levels is unexpected for CaO, given its ultrawide band gap and the doping bottlenecks known to exist in wide-band-gap materials.[14-18] This is compared to a range of VBM+0.57 eV to CBM−3.15 eV for BeO which has a band bap exceeding 10 eV.[10]

We now briefly discuss the feasibility of the NV-like centers in CaO recently predicted by Davidsson *et al.*[12] The most promising one of the NV-like centers consists of a bismuth substitution on the calcium site ($Bi_{Ca}$) adjacent to an oxygen vacancy ($V_O$) and is negatively charged, i.e., $[Bi_{Ca}V_O]^-$. This center requires to be made in a moderately to highly $n$-type doped CaO with the Fermi level close to CBM (more specifically, ~6 eV above the VBM; here the Fermi level is not referenced to the CBM due to uncertainty in the predicted CaO band gap in the work of Davidsson *et al.*[12]). Based on our predicted doping-limit energy range, the $[Bi_{Ca}V_O]^-$ can possibly be achieved, yet one needs to use a highly O-poor growth condition for CaO and find suitable $n$-type shallow dopants to move the Fermi level close to the CBM.

### C. Hydrogen impurities

In the above we have assessed the doping limits in CaO considering compensation by the intrinsic defects. We now examine how hydrogen impurities affect the doping limits. Hydrogen is a common impurity in oxides, often unintentionally incorporated during growth and processing.[19-22, 44, 45] For hydrogen impurities in CaO, we have studied H interstitial ($H_i$), H substitution on the O site ($H_O$), and H complexes with $V_{Ca}$ ($H + V_{Ca}$ and $2H + V_{Ca}$). Fig. 3 shows the formation energies of the hydrogen-related defects, under both O-poor and O-rich conditions and with the hydrogen chemical potential ($\mu_H$) set to the respective maximum allowed value (0 and −1.97 eV, respectively); see the supplementary material for details on how these $\mu_H$ values are obtained.

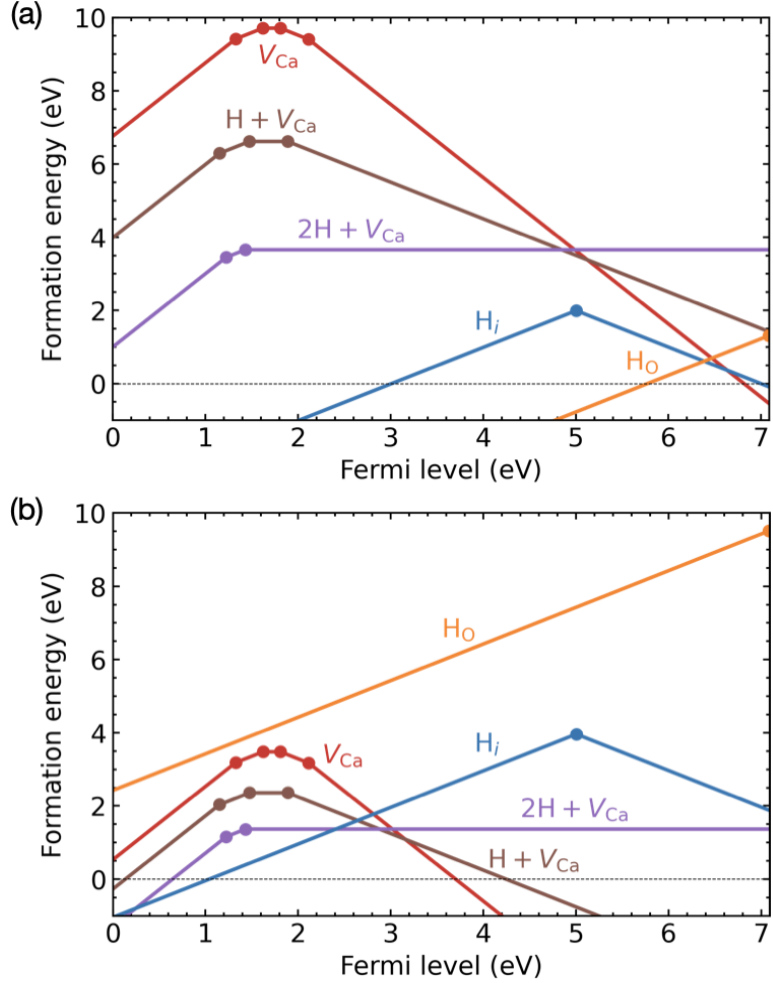

FIG. 3. Formation energy versus Fermi level for the $H_i$, $H_O$, $H + V_{Ca}$, and $2H + V_{Ca}$ in CaO, for (a) O-poor conditions and $\mu_H = 0$ eV and (b) O-rich conditions and $\mu_H = -1.97$ eV. The defect charge states are $H_i$: +, −; $H_O$: +, 0; $H + V_{Ca}$: 2+, +, 0, −; $2H + V_{Ca}$: 2+, +, 0. For comparison the formation energy of $V_{Ca}$ is also shown.

As in many crystalline materials,[31, 46-50] $H_i$ is an amphoteric center in CaO, with a deep (+/−) transition level (in the upper part of the band gap). The substitutional $H_O$ is a shallow donor and stable in the 1+ charge state. The shallow donor behavior of $H_O$ has been widely reported in other wide-band-gap oxides.[20, 31, 51-55] Compared to isolated $V_{Ca}$, the $H + V_{Ca}$ shows no 2− charge state and the $2H + V_{Ca}$ exhibits no negative charge states, which can be understood from hydrogen passivation of the $V_{Ca}$ dangling bonds. We find large (positive) binding energies of 2.11 eV for $[H + V_{Ca}]^-$ and 3.95 eV for $[2H + V_{Ca}]^0$ (see supplementary material for details), suggesting that the complexes would be stable if formed.

As shown in Fig. 3(a), under O-poor conditions and $\mu_H = 0$ eV (the H-rich limit), both $H_i$ and $H_O$ have low formation energy. Under such conditions, the $H_O$ will severely restrict the range of Fermi levels, by raising the p-type pinning limit to VBM+5.77 eV (compared to VBM+4.10 eV due to $V_O$). The $H_i$ will contribute to compensation of n-type doping. As shown in Fig. 3(b), under O-rich conditions and $\mu_H = -1.97$ eV (a very low $\mu_H$), the $H_O$ has high formation energy, while for $H_i$, the formation energy is low for Fermi level in the lower part of the band gap. Under such conditions, the p-type pinning limit will be raised by $H_i$ to VBM+1.04 eV (compared to VBM+0.53 eV due to $V_O$). We note that as long as $\mu_H$ is bound by the maximum allowed value for a given $\mu_O$, the hydrogen-related defects will not affect the *n*-type pinning limit (always decided by $V_{Ca}$).

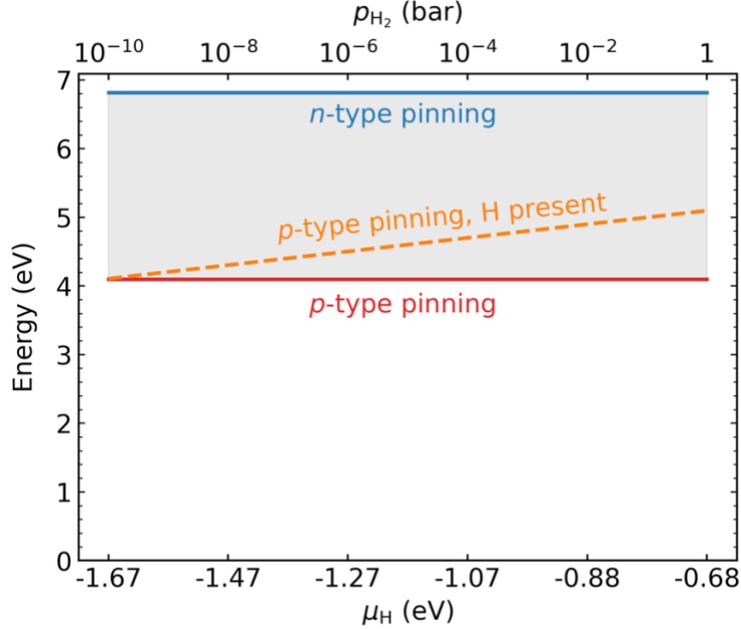

FIG. 4. Doping limits in O-poor CaO in the presence of hydrogen impurities. The $\mu_H$ values are obtained through equilibrium with H$_2$ gas at $T = 1000$ K and partial pressure indicated by the $p_{H_2}$ axis (on a logarithmic scale). The shaded area indicates the allowed range of Fermi levels, which gets narrowed in the presence of hydrogen impurities.

To define $\mu_H$ values that can reflect more realistic situations, we consider CaO in an H$_2$ atmosphere at 1000 K and different hydrogen partial pressure ($p_{H_2}$) ranging from $10^{-10}$ to 1 bar. The resulting $\mu_H$ values are between $-1.67$ and $-0.68$ eV [computed by Eq. S(2) in supplementary material]. Fixing $\mu_O$ to $-6.23$ eV (the O-poor limit), we plot in Fig. 4 the pinning energies as a function of $\mu_H$ ($p_{H_2}$). We see that as $\mu_H$ ($p_{H_2}$) increases, the *p*-type pinning limit shifts to higher energies in the band gap. The pressure $10^{-10}$ bar can be regarded as the lowest H$_2$ pressure that hydrogen can affect the doping limits in O-poor CaO. We could also consider the case $\mu_O = 0$ eV (the O-rich limit), but since in this case the maximum allowed $\mu_H$ ($-1.97$ eV) is below the $\mu_H$ at $10^{-10}$ bar, it is not discussed.

Finally, it is worth mentioning that besides acting as compensating donors, the shallow donor $H_O$, if present at high concentration in O-poor CaO, would raise the Fermi level and dope CaO *n*-type, as suggested by Fig. 3(a). Indeed, using the defect formation energies in Figs. 1(a) and 3(a) and assuming 1000 K synthesis followed by rapid quenching to room temperature, the Fermi level is calculated to be 0.21 eV below the CBM, resulting in a net electron concentration on the order of $10^{16}$ cm$^{-3}$. This suggests that hydrogen impurities could be exploited to prepare *n*-type CaO for hosting the NV-like center $[Bi_{Ca}V_O]^-$. This is assuming that hydrogen will not interact with the NV-like center.

## IV. Conclusions

In summary, based on first-principles hybrid-functional calculations, we find that $V_{Ca}$ and $V_O$ are the most common intrinsic defects in CaO, acting as compensating acceptors and donors, respectively. The $O_i$ is also prevailing under O-rich conditions and acts as a compensating donor. Due to compensation by these defects, O-poor conditions are required to dope CaO *n*-type, while O-rich conditions are required for *p*-type doping. For a given growth condition, the allowed range of Fermi levels is quite restricted. By adjusting growth conditions, intrinsic CaO can have an achievable range of Fermi levels between VBM+1.76 eV and CBM−1.73 eV (assuming high-temperature growth and rapid quenching to room temperature). Moving the Fermi level closer to the band edges will require shallow dopants but a wider Fermi-level range will be attainable: VBM+0.53 eV to CBM−0.27 eV. If hydrogen impurities are present, the hydrogen-related defects will cause additional limit to *p*-type doping, but on the other hand, the $H_O$ is a shallow donor which could contribute to *n*-type doping.

## ACKNOWLEDGMENTS

We acknowledge support from the U.S. Department of Energy, Office of Science, Basic Energy Sciences in Quantum Information Science under Award Number DE-SC0022289. This research used resources of the National Energy Research Scientific Computing Center, a DOE Office of Science User Facility supported by the Office of Science of the U.S. Department of Energy under Contract No. DE-AC02-05CH11231 using NERSC award BES-ERCAP0020966.## AUTHOR DECLARATIONS

### Conflict of interest

The authors have no conflicts to disclose.

## DATA AVAILABILITY

The data that support the findings of this study are available from the corresponding author upon reasonable request.